\documentclass[twoside,twocolumn,a4paper]{article}
\usepackage{graphicx,graphics}
\usepackage{psfrag}
\usepackage{amsmath,amssymb,amsfonts}
\usepackage{caption}
\usepackage{dcolumn}
\usepackage{bm}
\usepackage{epstopdf}
\usepackage[usenames,dvipsnames]{xcolor}
\usepackage[margin=1.7cm]{geometry}
\usepackage{authblk}

\newcommand{\ket}[1]{|#1\rangle}

\title{Coherent collective dynamics and  entanglement evolution of polar molecules on 1D lattices.}

\author[1]{Vanessa Olaya-Agudelo }
\author[1,2]{Karen Rodr\'{i}guez} 

\affil[1]{Departamento de F\'{i}sica, Universidad del Valle, A.A. 25360, Cali, Colombia}

\affil[2]{Centre for Bioinformatics and Photonics -- CiBioFi, Calle 13 No. 100-00, Edificio 320 No. 1069, Cali, Colombia}
\affil[ ]{ \small karem.c.rodriguez@correounivalle.edu.co}

\date{\today}
\begin{document}

\twocolumn[
  \begin{@twocolumnfalse}
    \maketitle
    \begin{abstract}
      We study a LiCs strongly-interacting molecular gas loaded into an one-dimensional optical lattice at quarter filling. The molecules are in the lowest electronic and vibrational state, $X^{1}\Sigma$ ($\nu=0$). Due to the large intermolecular distance and low filling, dipole-dipole interaction in the nearest-neighbor approximation governs the dynamics of the rotational excitations. For low DC electric field strengths, the full set of rotational levels $N=0,1$ must be taken into account, nevertheless, our calculations show that very weak fields act as field-selectors disclosing two- and three-level systems out of the original four-level one. The dynamics and the generated von Neumann entanglement entropy among the internal rotational states throughout the evolution are presented for low, moderate and strong fields . We observe a sharp and monotonous growth of the entanglement as the dynamics take place showing the potential of these molecular systems to be used in quantum information protocols. The numerical simulations are performed by means of the Time-Evolving Block Decimation algorithm based on the Matrix Product State formalism and the Susuki-Trotter decomposition.
\vspace{1cm}
\end{abstract}
\end{@twocolumnfalse}
]

\section{Introduction} \label{intro}

Ultracold and quantum degenerate molecular gases are currently a focus of interest within atomic physics, molecular physics and physical chemistry communities~\cite{Ni2010}.  The enormous control and high tunability of cold and ultracold molecular systems make them suitable to study a broad applications spectrum in diverse fields, from precision measurement and high-resolution spectroscopy \cite{Zelevinsky2008,DeMille2008} to ultracold chemistry \cite{Krems2005,Balakrishnan2016}, quantum information processing \cite{DeMille2002} and quantum computing~\cite{Micheli2006}.

Cold lattice gases, either atomic or molecular, are often proposed as quantum simulators mimicking, for instance, the Hubbard and Heisenberg \cite{Jaksch2005,Lewenstein2007,Rodriguez2010,Rodriguez2011} models, serving as a way to develop new experimental, theoretical and numerical tools. The production of ultracold ground-state molecules opens the path to derive effective many-body models for nonreactive molecules in tight traps like optical tweezers \cite{Kaufman2012,Thompson2013} and optical lattices \cite{Docaj2016}.  Arranging polar molecules in optical standing-wave trapping-potentials, new opportunities are expected due to the neighboring sites dipole-dipole interaction; in contrast, ground-state atomic lattice gases dominated by $1/R^6$ van der Waals forces exhibit only on-site interaction. Besides the long-range character of the dipolar interaction, it is interesting to explore its anisotropy which can be manipulated by applying external electric fields \cite{Rosario2007,Ni2010,Lameshko2013}.

The successful association of two pre-cooled atoms to form a molecule has been achieved using Feshbach resonances~\cite{Koehler2006} and photoassociation~\cite{Jones2006}. Magneto association performed by STIRAP has produced KRb molecules in their electronic and vibrational ground-state~\cite{Ni2008,Ospelkaus2008} and photoassociation close to a Feshbach resonance has produced LiCs molecules in their electronic and vibrational ground-state~\cite{Deiglmayr2008}.

It is of great interest to control the time-evolution among specific discrete quantum states, specifically, the array of long-range interacting polar molecules with rich internal structure form an excellent ground to comprehend complex systems relevant for quantum information processing and simulations~\cite{Covey2017}. For this porpoise, new theoretical and numerical tools are needed to understand the  many-body quantum systems such as rotational dynamics present in state of the art experiments.

Therefore, in the present work we analyze the coherent population transfer of $X^{1}\Sigma$ molecules due to the dipole-dipole interaction in their ground and first excited rotational levels as a function of an external electric field.  Pursuing this aim, the molecular gas is loaded onto an one-dimensional optical lattice and the gas is prepared in the strongly-correlated Mott-insulator regime.

Recently, the use of quantum information tools in several communities has been growth~\cite{Horodecki2009,Eisert2010}. Particularly, in the field of the many-body strongly-correlated quantum-systems~\cite{Amico2008,Islam2015}, these tools allow us to analyze nonlocal information and therefore a suitable mechanism to study quantum phases and phase transitions~\cite{mishmash2016,matsuura2016,He2017}. In a pure bipartite state, having measured only part of the system (a subsystem) it is possible to have information of the whole quantum state if there is entanglement among the parts~\cite{Calabrese2009}. In this paper, we use the von Neumann entropy associated to the reduced density matrix to provide the entanglement entropy of the measured subsystem.

This paper is organized as follows: in sect. \ref{model} we describe both the model of the molecular lattice system and the implemented numerical method. Section \ref{dynamics} is devoted to molecular excitations dynamics in presence of an external DC electric field. The entanglement generated throughout the evolution of the excitations is presented in sect. \ref{ent}. Finally, sect. \ref{summary} concludes.

\section{Model and methods} \label{model}

\subsection{Molecular Hamiltonian in the Mott-insulator regime} \label{hamiltonian}

We consider an optical lattice filled with $X^1 \Sigma$ LiCs mole\-cules which are in the electronic and vibrational ground states. The lattice holds one molecule per site at a distance of 400 nm, hence, tunneling between lattice sites is completely suppressed \cite{Herrera2010}. Considering this low lattice filling, the on-site Hubbard interaction can be neglected as well. Thus, the interaction between the molecules is solely determined by the long-range anisotropic dipole-dipole interaction, creating quasi-partic\-les responsible for excitation transfer along the lattice, so-called Frenkel excitons~\cite{Agranovich2008}. The only two states involved in the dynamics are the ground and the first excited rotational states. The Hamiltonian is given by~\cite{Herrera2010},

\begin{equation} \label{Excitation}
\hat H=\sum_{n=1}^{N_{\text{mol}}} \left(B_e\hat N_n^2-{\bold d}_n\cdot{\bold E}\right)+\frac{1}{2}\sum_{n=1}^{N_{\text{mol}}}\sum_{m\ne n}^{N_{\text{mol}}} \hat V_{dd}({\bold r}_n-{\bold r}_m),
\end{equation}

where ${\bold r}_n$ is the position of the $n$-th lattice site, $B_e$ is the rotational constant, ${\bold d}_n$ is the electric dipole operator of individual molecules, ${\bold E}$ is the constant electric field, $\hat{N}_n$ is the rotational angular momentum operator of the molecule at the site $n$ and $\hat{V}_{dd}$ is the dipole-dipole interaction between molecules in different lattice sites. $N_{mol}$ is the total number of molecules.

The local rotational Hamiltonian, $\sum_n B_e\hat N_n^2$, corresponds to the crystal energy operator (rotational energy) without intermolecular interactions. In absence of a static electric field, the rotational ground state of the given diatomic molecule has a rotational angular momentum $N=0$, and the first excited rotational state $N=1$ is triply degenerate. In the presence of a field  ${\bold E}$,  the Stark effect is accounted for, the degeneracy of $N=1$ is lifted and splits into two sub-energy levels corresponding to the projections, $M=0$ and $M=\pm 1$, into the electric field quantization axis.  For strong fields beyond a critical value (see reference \cite{Lincoln2009} for critical field theory), the system reduces to a two-level problem considering exclusively $M=0$ projections. In contrast, for a weak static electric field all four levels must be considered~\cite{Lincoln2009}, this is the scenario considered in the present work.

The electric field acts only on the internal coordinates of single molecules, and couples different local molecular states, this coupling depends on the electric field strength.

The dipole-dipole operator, $\hat V_{dd}({\bold r}_n-{\bold r}_m)$, involves two different sites and mixes two different molecular states. In fact this interaction determines the dynamics of the system and their features. In the Molecular Fixed Frame (MFF) this operator takes the form,

\begin{equation} \label{VddMFF}
\hat{V}_{dd}(\textbf{r})=\frac{1}{{r}^{3}}[\hat{d}_{n}\cdot\hat{d}_{m}-3(\hat{d}_{n}\cdot\hat{e})(\hat{e}\cdot\hat{d}_{m})].
\end{equation}

Here, $\textbf{r}={\bold r}_n-{\bold r}_m$ is the intermolecular vector, $\hat{e}=\frac{\bold{r}}{r}$ is the unit vector in the direction $\bold{r}$ and $\hat{d}_{m}, \hat{d}_{n}$ are the dipole moment operators of the molecules at the sites $m$ and $n$, respectively.

In order to study the interplay of the local interaction with $\textbf{E}$ and the non-local dipole-dipole interaction, it is needed to represent the operator (\ref{VddMFF}) in the Laboratory Fixed Frame (LFF),

\begin{equation} \label{VddLFF}
\hat{V}_{dd}(\textbf{r})=-\frac{\sqrt{6}}{{r}^{3}}\sum_{q=-2}^{2}{-1}^{q}C_{-q}^{2}(\textbf{r})[\hat{d}_{n}\otimes\hat{d}_{m}]_{q}^{(2)}.
\end{equation}

Here $C_{-q}^{2}(\textbf{r})=\sqrt{\frac{4\pi}{5}}Y^{(2)}_{-q}(\textbf{r})$ are the reduced spherical harmonics which describe the movement of the intermolecular axis in the LFF. The one-dimensional confinement of the lattice reduces this coefficient to a simple factor. The operators on the right hand side in Eq. (\ref{VddLFF}) contain the information about the strength of the involved molecular dipole moments. The term can be seen as a second rank tensor product of two tensors of rank one $\hat{T}_{i}^{1}(\hat{d}_{s})$ \cite{Zare} which represents the dipole moment operator of the molecule located in the site $s \in \{n,m\}$. As a result, we express $[\hat{d}_{n}\otimes\hat{d}_{m}]_{q}^{(2)}$ as $\sum_{p}\langle 1 p, 1 q-p | 2 q \rangle\hat{T}_{p}^{1}(\hat{d}_{n})\hat{T}_{q-p}^{1}(\hat{d}_{m})$.

Using the second quantization, the Hamiltonian (\ref{Excitation}) can be written following Ref.~\cite{Agranovich2008} as

\begin{equation} \label{finalH}
\hat{H}_{\text{exc}}=\sum_{n}E_{0}\hat{P}_{n,M}^{\dagger}\hat{P}_{n,M}+\sum_{n,m \neq n}\sum_{M,M'}F_{n,m}^{M,M'}\hat{P}_{n,M}^{\dagger}\hat{P}_{m,M'}.
\end{equation}

Here $E_{0}$ is the energy difference between the excited and ground rotational states, $\hat{P}_{n,M}^{\dagger}$ $(\hat{P}_{m,M'})$ creates (annihilates) a molecule in the site $n$ $(m)$ and level $M$ $(M')$. Specifically, the operators act as $\hat{P}_{n,M}^{\dagger} |g ,0\rangle _n = |e,M \rangle _n$ and $\hat{P}_{m,M'} |e,M' \rangle _m = |g \rangle _n$, where $|g,0 \rangle _n$ and $|e,M \rangle _n$ are the field-dressed ground and excited states, respectively. The coefficients $F_{n,m}^{M,M'}=\langle e_{n,M'} g_{m,0} | \hat{V}_{dd} | g_{n,0} e_{m,M} \rangle $ provide the molecular selection ru\-les, which determine the restricted phase space accessible at low-energies. Here, we use a DC electric field perpendicular to the molecular arrangement leading to the selection rule $\Delta (M_{n}+M_{m}) = 0, \pm 2$ involving molecules on sites $n$ and $m$~\cite{Herrera2010}. Finally, working with molecules separated distances greater than $100$ nm makes the intermolecular interaction considerably smaller compared to the rotational spectrum energy, so it is enough to account only for the nearest-neighbor interaction.

The Hamiltonian (\ref{finalH}) allows us to describe the behavior of the excitations as a function of the external electric field and study the coherent population transfer between the two involved rotational molecular levels.

We consider a lattice array with $L$ sites, such bipartite system can be separated by applying the Schmidt decomposition~\cite{Schmidt1907}. Each state can be written as $\ket{\Psi} = \sum_{\alpha_L =1}^{N_B} \lambda_\alpha \ket{\alpha_B} \otimes\ket{\alpha_{E}}$, where $\{\ket{\alpha_B}\}$ (on one subsystem with $N_B$ Hilbert dimension) and $\{\ket{\alpha_E}\}$ (on the other with $N_E$ Hilbert dimension) are two orthonormal basis sets belonging to the respective Hilbert subspace, each with at most $N_B$ basis elements, here $N_E>N_{B}$. The coefficients $\lambda_{\alpha}$ are non-negative real numbers satisfying $\sum_{\alpha}^{\chi}\lambda_{\alpha}^2=1$, known as the {\it Schmidt coefficients} and $\alpha$ runs in the reduced Hilbert subspace. In the numerical implementation the sum is truncated and only $\chi$ states are kept.

In a pure state there is only one density matrix eigenvalue. But the reduced density matrix for a subsystem $\hat\rho_B=\sum_\alpha\lambda_\alpha^2\ket{\alpha_B}\ket{\alpha_B}$ represents a mixed state where the entanglement entropy can be measured using the von Neumann entropy,

\begin{equation} \label{entropy}
S=-\sum_\alpha \lambda_\alpha^2 \ln (\lambda_\alpha^2).
\end{equation}

\subsection{Time evolution}

Numerical calculations reveal important aspects in the ultracold interacting molecular behavior.  In low dimensions, quantum fluctuations are enhanced and therefore many-body quantum systems, in general, are not susceptible to the standard mean field or perturbation expansion methods so one has to turn to non-perturbative techniques and numerical methods. Here, we use the Matrix Product State (MPS) ansatz~\cite{Vidal2003,Vidal2004,Verstraete2004,Verstraete-Porras2004} to efficiently approximate the quantum states. This formalism is highly recommended when the amount of entanglement in the system is limited, reducing the calculation complexity from the original exponential to an algebraic growth of the Hilbert space. Based on the MPS representation, we use the Time-Evolving Block Decimation (TEBD) algorithm~\cite{Vidal2004} to numerically analyze the molecular quantum-chain dynamics, paying special attention to population transfer and quantum entanglement. The TEBD takes advantage of the fact that the Hamiltonian can be written as the sum over {\it even} and {\it odd} sites and then the time evolution operator is approximated using the Trotter-Suzuki expansion formula~\cite{Suzuki1990}. Since the entanglement of the system grows with time, keeping the matrices of the same size means to loose information throughout the evolution. In order to avoid so, one has to adapt the value of the entanglement parameter ($\chi$) every time-step to keep the state as accurate as possible~\cite{Daley2004}.

The approximations discussed in sect.~\ref{hamiltonian} allow us to use this numerical technique for calculating the time evolution since the dipole-dipole interaction couples only nearest-neighbors~\cite{Rabl2007}. In the following, we consider chains with $L=16$ sites and initial entanglement parameter $\chi_{0}=8$ since the evolution starts from an array of disentangled individual molecules. Towards the end, the entanglement grows and we reach $\chi_{\text{final}}=100$ in average after 660 time-steps.

\section{Dynamics of the molecular excitations}\label{dynamics}

In this section, we first analyze the dynamics of the molecular excitations in the presence of a static and perpendicular electric field. Several scenarios are regarded depending on the field strength and the initial condition.

We work with LiCs alkali molecules, whose dipole moment is $d_0 = 5,523$ De~\cite{Aymar2005}, the rotational constant is $B_e = 5,816 $GHz and the critical electric field is $E_{\text{cri}} = 2100 $V/cm \cite{Lincoln2009}. Initially all lattice sites are prepared in the same state ($|\Psi_0\rangle$) which consists of either a single rotational level or a superposition of them. The choices are motivated by two ideas; first, the possibility to reach these single-molecule states after two-atoms association mechanisms. And second, the access to the strongly-correlated regime with dimers with tools developed in atomic physics field~\cite{Jaksch2005,Volz2006}.

As the system evolves the rotational excitations exchange and spread throughout the lattice following the selection rules. Hence, we locally measure the rotational state of each molecule, sum over all sites and divide by the total initial amount of molecules. Therefore, the figures presented in this section show the percent population of each single dressed state as a function of time. The time is given in unities of $(\mu^2/d^3)^{-1}$ which for our case is of the order of $\sim 9\times 10^4$ GHz$^{-1}$.

\begin{figure}
\psfrag{Population (\%)}{\hspace{-0.3cm}\footnotesize Population $(\%)$}
\psfrag{Time}{\hspace{-0.6cm}\footnotesize Time $(\mu^2/d^3)^{-1}$}
\psfrag{0,0}{{\hspace{-0.3cm}\scriptsize  $|0,0\rangle$}}
\psfrag{1,-1}{{\hspace{-0.3cm}\scriptsize  $|1,\text{-}1\rangle$}}
\psfrag{1,0}{{\hspace{-0.3cm}\scriptsize $|1,0\rangle$}}
\psfrag{1,1}{{\hspace{-0.3cm}\scriptsize $|1,1\rangle$}}
\psfrag{(a)}{{\hspace{-1.1cm}\scriptsize (a) $|0,0\rangle$}}
\psfrag{(b)}{{\hspace{-1.1cm}\scriptsize (b) $|1,0\rangle$}}
\psfrag{(c)}{{\hspace{-1.1cm}\scriptsize (c) $|1,1\rangle$}}
\psfrag{(d)}{{\hspace{-1.1cm}\scriptsize (d) $|\Psi_1\rangle$}}
\psfrag{(e)}{{\hspace{-1.1cm}\scriptsize (e) $|\Psi_2\rangle$}}
\psfrag{(f)}{{\hspace{-1.1cm}\scriptsize (f) $|\Psi_3\rangle$}}
\includegraphics[width=1\linewidth]{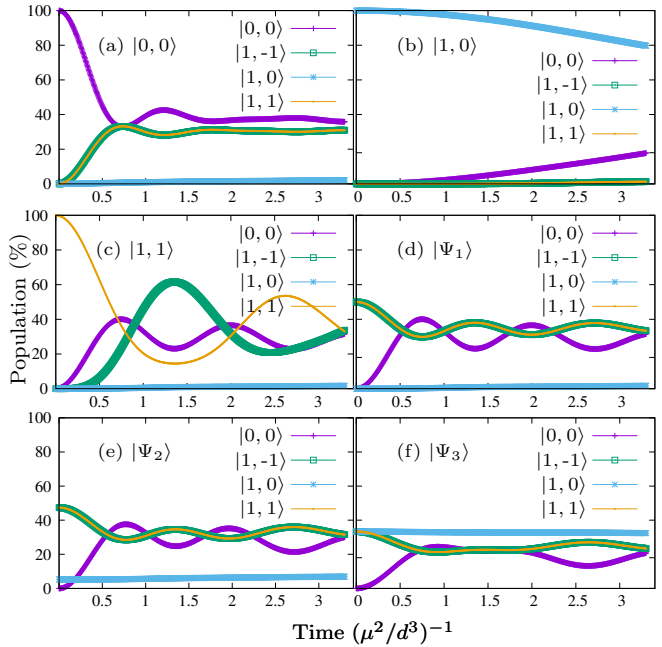}
\caption{Coherent population transfer among the rotational states in presence of a weak DC electric field of $E=50$ V/cm. All molecules are prepared in the same Fock state (a) $|0,0\rangle$, (b) $|1,0\rangle$ and (c) $|1,1\rangle$ and the superposition (d) $|\Psi_1\rangle=\frac{1}{\sqrt{2}}\big(|1,\text{-}1 \rangle + |1,1 \rangle\big)$, (e) $|\Psi_2\rangle=\tfrac{1}{\sqrt{19}}\big(3|1,\text{-}1\rangle + |1,0 \rangle+ 3|1,1 \rangle\big)$ and (f) $|\Psi_3\rangle=\frac{1}{\sqrt{3}}\big(|1,\text{-}1 \rangle + | 1,0 \rangle + | 1,1 \rangle\big)$. See color figures online.\label{E50}}
\end{figure}

\subsection {Weak field: Effective two- and three-level dynamics}

Let us start by analyzing the case where a weak field is applied, namely $E=50 $ V/cm.
For a very low external field, the coupling between $| 0,0 \rangle$ and $| 1,0 \rangle$ is weak. This physical fact is presented in Fig.~\ref{E50}, starting from the $| 0,0 \rangle$ state (see Fig. \ref{E50}(a)), the population rapidly starts to decrease and head towards the two degenerate states $|1,-1 \rangle$ and $| 1,1 \rangle$, while $|1,0 \rangle$ state remains unchanged up to $t=0.5$ and slightly, almost imperceptible, rises its population. 
On the contrary, when all the population starts in the $|1,0 \rangle$ state, the excluded states from the exchange are $| 1,-1 \rangle$ and $| 1,1 \rangle$ (see Fig. \ref{E50}(b)). The system behaves as a {\it two-level system}, besides the initial population does not decrease dramatically as in  \ref{E50}(a), but slowly go to the coupled state $|0,0 \rangle$. Initiating from $|1,1 \rangle$  the population first go to the ground state due to the selection rules and later on the state $| 1,-1 \rangle$ begins to be populated. It can be seen how $| 1,-1 \rangle$ takes a while before enters into the dynamics and afterwards $| 0,0 \rangle$, $| 1,-1 \rangle$ and $| 1,1 \rangle$ coherently oscillate.

Otherwise, Figs.~\ref{E50}(d), (e) and (f) show different initial population configurations, the evolutions start from linear superpositions of $| 1,-1 \rangle$, $| 1,0 \rangle$ and $| 1,1 \rangle$ states with different contributions.  For these three cases the systems behave effectively as {\it three-level systems} involving only the states: $\{|0,0\rangle, |1,-1\rangle, |1,1\rangle\}$, excluding the contribution of the $|1,0\rangle$ state, which keeps the population practically constant for the whole evolution. The same behavior applies for Figs.~\ref{E50}(a) and (c).

Hence, this low DC field strength allows us, depending on the initial condition, to switch from a four-level system to a two- or a three-level one.

\subsection {Moderate field}
\begin{figure}[t]
\psfrag{Population (\%)}{\hspace{-0.4cm} \footnotesize Population $(\%)$}
\psfrag{Time}{\hspace{-0.6cm}\footnotesize Time $(\mu^2/d^3)^{-1}$}
\psfrag{0,0}{{\hspace{-0.3cm}\scriptsize  $|0,0\rangle$}}
\psfrag{1,-1}{{\hspace{-0.3cm}\scriptsize  $|1,\text{-}1\rangle$}}
\psfrag{1,0}{{\hspace{-0.3cm}\scriptsize $|1,0\rangle$}}
\psfrag{1,1}{{\hspace{-0.3cm}\scriptsize $|1,1\rangle$}}
\psfrag{(a)}{{\hspace{-1.1cm}\scriptsize (a) $|0,0\rangle$}}
\psfrag{(b)}{{\hspace{-1.1cm}\scriptsize (b) $|1,0\rangle$}}
\psfrag{(c)}{{\hspace{-1.1cm}\scriptsize (c) $|1,1\rangle$}}
\psfrag{(d)}{{\hspace{-1.1cm}\scriptsize (d) $|\Psi_1\rangle$}}
\psfrag{(e)}{{\hspace{-1.1cm}\scriptsize (e) $|\Psi_2\rangle$}}
\psfrag{(f)}{{\hspace{-1.1cm}\scriptsize (f) $|\Psi_3\rangle$}}
\includegraphics[width=1\linewidth]{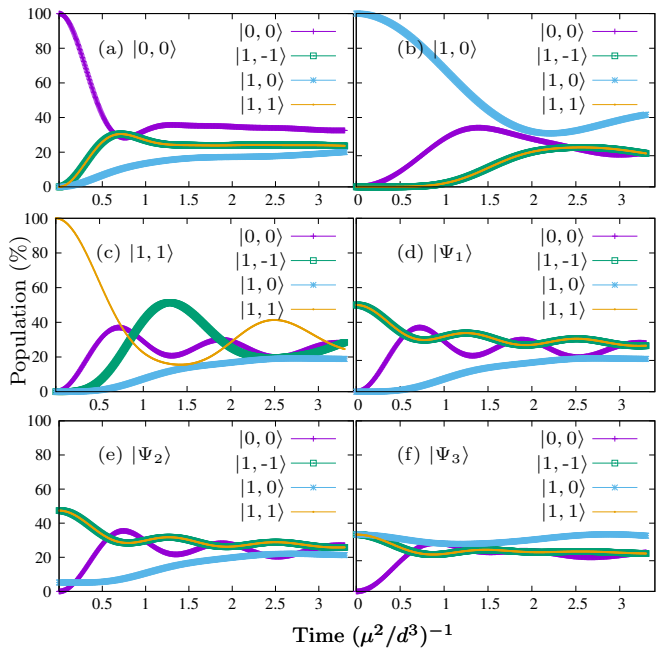}
\caption{Coherent evolution starting all molecules from the initial conditions defined in Fig.~\ref{E50} and using a moderate DC electric field strength of $E=200$ V/cm. See color figures online.\label{E200}}
\end{figure}

As the field strength rises, all four states are involved in the dynamics. Here we use $E=200$ V/cm and the coherent collective evolutions are shown in Fig. \ref{E200}. Starting from the state $|0,0 \rangle$ (Fig.~\ref{E200}(a)) or state $|1,0 \rangle$ (Fig.~\ref{E200}(b)), the difference directly regards the selection rules. In the case of \ref{E200}(a) the excitations migrate directly to $N=1$, whereas in the case of \ref{E200}(b) the excitations must first go to the ground state and thereafter  populate the $M=\pm 1$ states. Hence, the states $|1, \pm 1 \rangle$ remain unpopulated for a longer period of time. 

The coupling between $M=0$ projections gets stronger than the latter case ($E=50$ V/cm), so in Fig.~\ref{E200}(a) the population of $|1,0 \rangle$ state heightens up to $20\%$ by the end of the time evolution, and the smooth coherent oscillations between $| 0,0 \rangle$ and $| 1,\pm 1 \rangle$ observed in Fig.~\ref{E50}(a) are weakened even more for $E=200$ V/cm. On top of that, after $t=1.5$ the populations in $|1, \pm 1 \rangle$  level off to a stationary regime getting 25\% of the total population each. From this point on the dynamics reach again an effective {\it two-level like system}.
In Fig.~\ref{E200}(b) the interaction between ground $N=0$ and exited states $N=1$ increases and the exchanges is higher than the case with lower electric field \ref{E50}(b), and slight oscillations among the rotational levels appear.

When the $|1,1 \rangle$ state is taken to be the initial state (similar analysis applies for $|1,-1 \rangle$), see Fig. \ref{E200}(c), the population rapidly drops into $|0,0 \rangle$ state, then goes into $|1,-1 \rangle$ and finally the $|1,0 \rangle$ enters in the dynamics. The field dressed states with $M = \pm 1$ have the same energy, hence, they have the same probability to be populated, being the reason why $|1,-1 \rangle$ state comes first in the dynamics than $|1,0 \rangle$ state. High-amplitude coherent oscillations are presented for this initial condition and the considered field strengths.
 
For the case shown in Fig.~\ref{E200}(d), the evolution starts from a superposition state with equal populations (symmetric Bell-like state) using $|1,\pm 1 \rangle$ instead of a Fock state. As it is observed, the rotational ground state becomes quickly populated while $|1,0 \rangle$ rises late and slowly in the dynamics. The latter levels out reaching a steady regime whilst the populations of the other three states oscillate coherently, regardless of the initial Fock state (c) or Bell superposition (d). 

As long as $|1,\pm 1 \rangle$ dressed states are initially populated, the $M=0$ projection of the $N=1$ state behaves in the same manner matching actually their time-scales. This observation still applies even when the $|1,0 \rangle$ state is allowed to be populated with almost 10$\%$ of the total population, see Fig. \ref{E200}(e) (calculations where performed also with 20$\%$ of the population delivering the same result) and the behavior for this state persists: remaining constant at the initial population for a while, and then rising slowly until it levels out.

Finally we explore the case when all states in the rotational excited state $N=1$ have the same initial weight achieving the evolution presented in Fig. \ref{E200}(f). In this case, the rotational level $|1,0\rangle$ has the largest probability of being populated as the evolution takes place, while the populations in the ground state and projections $M=\pm 1$ tend to a quasi-steady regime sharing the same probability percentage.
Between $t=0.5$ and $t=1$, the ground state reaches its maximum, almost when the population in $| 1,0 \rangle$ stars to rise, see Figs.~\ref{E200}(c), (d) and (f). Therefore, for the ground state to get populated a elapsed time around $(\mu^2/d^3)^{-1}$ is needed, afterwards a redistribution to the other leves is performed according to selection rules and weight probabilities.

\subsection {Strong field}
\begin{figure}[t]
\psfrag{Population (\%)}{\hspace{-0.3cm}\footnotesize Population $(\%)$}
\psfrag{Time}{\hspace{-0.6cm}\footnotesize Time $(\mu^2/d^3)^{-1}$}
\psfrag{0,0}{{\hspace{-0.3cm}\scriptsize  $|0,0\rangle$}}
\psfrag{1,-1}{{\hspace{-0.3cm}\scriptsize  $|1,\text{-}1\rangle$}}
\psfrag{1,0}{{\hspace{-0.3cm}\scriptsize $|1,0\rangle$}}
\psfrag{1,1}{{\hspace{-0.3cm}\scriptsize $|1,1\rangle$}}
\psfrag{(a)}{{\hspace{-1.1cm}\scriptsize (a) $|0,0\rangle$}}
\psfrag{(b)}{{\hspace{-1.1cm}\scriptsize (b) $|1,0\rangle$}}
\psfrag{(c)}{{\hspace{-1.1cm}\scriptsize (c) $|1,1\rangle$}}
\psfrag{(d)}{{\hspace{-1.1cm}\scriptsize (d) $|\Psi_1\rangle$}}
\psfrag{(e)}{{\hspace{-1.1cm}\scriptsize (e) $|\Psi_2\rangle$}}
\psfrag{(f)}{{\hspace{-1.1cm}\scriptsize (f) $|\Psi_3\rangle$}}
\includegraphics[width=1\linewidth]{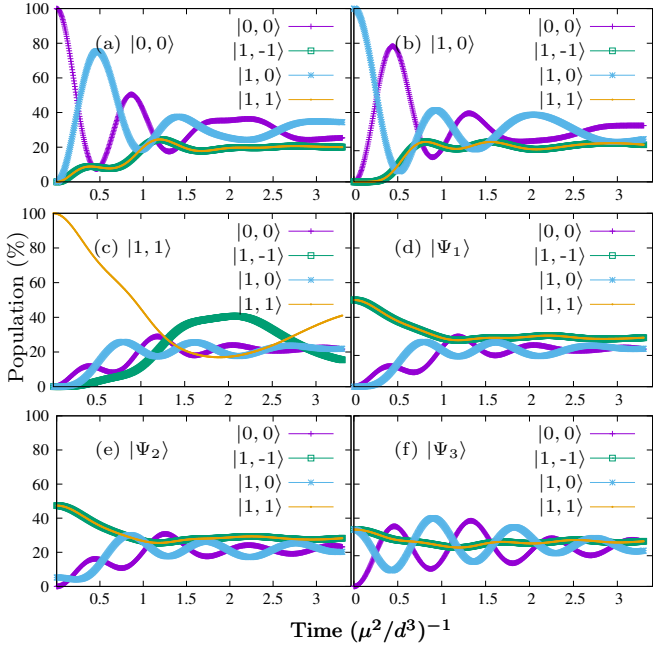}
\caption{Rotational levels dynamics with all molecules prepared in the initial conditions defined in Fig.~\ref{E50} using a strong DC electric field of $E=1000$ V/cm. See color figures online.\label{E1000}}
\end{figure}

Let us now increase the electric field to $E=1000$ V/cm which is still small compared to the rotational molecular energy. In this scenario, the coupling between the states with $M=0$ projection gets stronger, therefore the dynamics tend to be dominated by these two states, while the $M=\pm 1$ projections behave smoothly and the evolutions for several initial states are presented in Fig.~\ref{E1000}.

Let us consider two initial cases, when all molecules are prepared in the rotational ground state (Fig. \ref{E1000}(a)) and in the first rotational excited state with $M=0$ (Fig. \ref{E1000}(b)). In both cases the initial state drops dramatically its population  as much as the coupled state with the same projection rises and the populations oscillate strongly between the $m=0$ projections. Thereafter, these two states exchange coherently most of the chain population while the  $|1, \pm 1 \rangle$ dressed states starting from zero have an upward trend, with a slight decrease, stabilizing towards the end. For the evolution \ref{E1000}(b) $M=\pm 1$ states remain unpopulated for a longer period of time climbing up when the populations in $|0,0\rangle$ and $|1,0\rangle$ match each other, leveling out although swinging slightly until the end of the evolution.

Taking $|1,1 \rangle$ as the initial state leads to a different time evolution (see Fig.~\ref{E1000}(c)). According to the selection rules, molecules go first to the ground state and since the $M=0$-coupling becomes stronger due to the hight field, the $|1,0 \rangle$ state has a larger probability to be populated. Hence, in contrast of the previous case (Fig.~\ref{E200}(c)), the level population order changes and the ground state enters first in the dynamic, then the $| 1,0 \rangle$ state takes advantage over $| 1,-1 \rangle$ that comes latter. Once all levels get populated the states with $m=0$ oscillate coherently, while the $|1, \pm 1 \rangle$ states are struggling to equate their populations. The maximum of  $| 1,- 1 \rangle$ is reached approximately at $t\sim 2$.

When launching the evolution with each molecule prepared in a Bell state made up of $|1,\pm 1\rangle$ (see Fig.~\ref{E1000}(d)), the population in both levels decreases continuously during the first time unity and remains stable around 30\% throughout the rest of the evolution. Meanwhile, the $|0,0 \rangle$ and $|1,0 \rangle$ states behave in the same manner as in Fig.~\ref{E1000}(c). Modifying the initial state by increasing the population in $|1,0 \rangle$ does not change the general demeanor of the evolution of the $M=\pm 1$ projections as shown in Fig.~\ref{E1000}(e). This is true even if the initial population in $|1,0 \rangle$ equals $M=\pm 1$ as observed in Fig.~\ref{E1000}(f). In the latter, we analyze the initial state with the molecules in a linear combination of equally probable $N=1$ states. Here, one can see that after decreasing the population in $M=\pm 1$ level almost to a constant value whilst both $M=0$ projections oscillate coherently damping down towards the end of the evolution.

Moderate and strong field strengths lead, regardless the initial conditions, the evolutions to a population distribution between $20 \%$ and $40 \%$. Hence, for longer period of times, out of our calculations scope, we infer that the population  distribution in the steady state is expected to land in the same range. 

\section{Entanglement evolution.} \label{ent}

\begin{figure}[t]
\psfrag{Time}{\hspace{-0.6cm}\footnotesize Time $(\mu^2/d^3)^{-1}$}
\psfrag{0s}{{\hspace{-0.21cm}\scriptsize $|0,0\rangle$}}
\psfrag{0t}{{\hspace{-0.228cm}\scriptsize $|1,0\rangle$}}
\psfrag{1}{{\hspace{-0.31cm}\scriptsize $|1,1\rangle$}}
\psfrag{111}{{\hspace{-0.1cm}\scriptsize $|\Psi_1\rangle$}}
\psfrag{313}{{\hspace{-0.1cm}\scriptsize $|\Psi_2\rangle$}}
\psfrag{Entanglement}{\hspace{-0.6cm}\scriptsize Entanglement entropy}
\includegraphics[width=\linewidth]{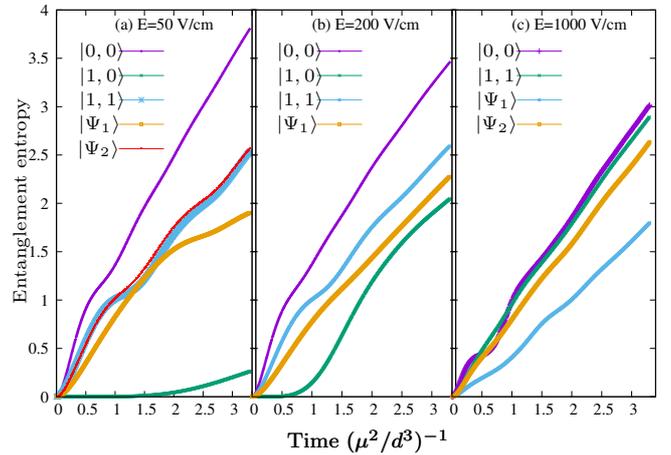}
\caption{Von Neumann entanglement entropy generated in the dynamics presented in Figs.~\ref{E50}, \ref{E200} and \ref{E1000} using as initial states: the Fock states $\{|0,0\rangle,|1,0\rangle,|1,1\rangle\}$ and the superpositions $\{|\Psi_1\rangle,|\Psi_2\rangle\}$ with $|\Psi_1\rangle=\tfrac{1}{\sqrt{3}}\big(|1,\text{-}1 \rangle + | 1,0 \rangle+ | 1,1 \rangle\big)$ and $|\Psi_2\rangle=\tfrac{1}{\sqrt{19}}\big(3| 1,\text{-}1 \rangle + |1,0 \rangle + 3|1,1 \rangle\big)$. Color figures online.\label{Ent}}
\end{figure}

In Sec.~\ref{dynamics} the evolution of the rotational levels has been presented depending on the initial state. Now, we focus on how the von Neumann entanglement entropy of the system grows as a function of time for three specific DC field strengths and several initial conditions. The corresponding results are shown in Fig.~\ref{Ent}.

At $E=50$ V/cm, see Fig.~\ref{Ent}(a), the system develops the highest but also the lowest entanglement observed in our numerical calculations. 
 Starting all molecules in the rotational ground-state turns out to be the initial condition that most entangles the system (lila curve), this is due to the fact that this state interacts with all the excited states, backed up by the selection rules established by the dipole-dipole interaction. Let us recall that for this electric field strength the state $|1,0\rangle$ is excluded while the populations in $M=\pm 1$ states increase until the three involved states level out reaching a quasi-stationary regime. Therefore, it is expected that initiating from $|1,0\rangle$ (green curve) the evolution leads to the most disentangled dynamic and this is exactly what we observed.
 When the majority contribution to the initial state comes from $ M= \pm 1$ states (red and blue curves), independently if it is a Fock or superposition state, due to the selection rules these levels have access only to the ground state, thus the system develops similar entanglement throughout the evolution.
 Even when the initial state, $|\Psi_2\rangle=\tfrac{1}{\sqrt{19}}\big(3| 1,\text{-}1 \rangle + |1,0 \rangle + 3|1,1 \rangle)$, which has a partial contribution of $| 1,0 \rangle$, the greatest contribution  to the dynamics comes from $M = \pm 1$ states, which determine the entanglement evolution.
A different scenario is presented when launching the evolution from an equally probable superposition of the excited rotational states (orange curve), where the entanglement grows steadily changing the slope to a lower rate at two time-units. This is because such superposition connects exclusively to the ground state by the selection rules.

As the field strength is increased to $E=200$ V/cm, see Fig.~\ref{Ent}(b), all evolutions rapidly entangle except the one when taking $|1,0\rangle$ as initial condition (green curve). In the dynamic this initial state mostly interacts with $ | 0,0 \rangle$ state, while the interaction with $m=\pm1$ is through of  $| 0,0 \rangle$. Notice that initiating from $| 1,1 \rangle$ and the superposition $|\Psi_1\rangle=\tfrac{1}{\sqrt{3}}\big(|1,\text{-}1 \rangle + | 1,0 \rangle + | 1,1 \rangle\big)$  lead to two different dynamics (Figs.~\ref{E200}(c) and (f)), but similar entanglement evolution, since both states transfer their population directly to the ground state. Let us point out that the coupling between $| 1,0 \rangle$ and $| 0,0 \rangle$ gets enhanced as the external electric field increases, hence it differs from the interaction among the levels $| 1, \pm 1 \rangle$ and $| 0, 0 \rangle$.

 Finally in the case of $E=1000$ V/cm, see Fig.~\ref{Ent}(c), most of the curves coincide except for the evolution of the superposition with equally probable $N=1$ states whose evolution does not match the rest although keeps the same behavior below the others. For this electric field the coupling between $M=0$ states gets even stronger, and $M=\pm 1$ states gets weaker, then for superposition states (blue and orange curves) that include equally probable $M=\pm1$ states, their entanglement evolution slow down.

 Overall, it is observed that the weaker the field the higher the entanglement rises for the ground state, this is due to its connection with all $N=1$ states. As the electric field increases, the coupling between $M=0$ levels is favored over $M=\pm1$ states and the amount of states involved throughout the evolution decreases and the entanglement in the chain decreases as well.

Let us note that at some point all curves must stop the growing tendency and level out which is not presented in our calculations. This is due to the fact that the entanglement for finite size systems do not grow indefinitely rather reach a logarithmic behavior in 1D, and our calculations show only an early evolution of the molecular chain.

\section{Conclusions} \label{summary}

We studied a LiCs molecular gas loaded on an 1D optical lattice prepared in the strongly-correlated Mott-insulator regime with one molecule per lattice site.  The molecules are in the lowest electronic and vibrational state and the coherent population transfer between the internal rotational states as a function of an external DC electric field is analyzed.  Due to the large intermolecular lattice distance and low filling, the hopping between nearest-neighbors and the on-site Hubbard interaction are suppressed, hence, the dipole-dipole interaction in the near\-est-neighbor approximation solely governs the dynamics of the excitations. For field strengths lower than the critical field the full set of rotational levels must be taken into account and the numerical simulations, here reported, confirmed this fact. Coherent transfer of population among the field-dressed states along the lattice are shown for several electric field intensities and initial conditions. Our numerical simulations show that weak fields have the potential to peak either a two- or three-level system out of the original four-level one, therefore, the field acts as field-selector for the molecular system, simplifying the complexity for both numerical and analytical calculations. Even more, the excluded states can be thought of as dark states and can be used to store unperturbed information throughout the dynamics regarding the initial condition.

Numerical simulations, performed with three field stre\-ngths, showed that molecular excitations become entangled as the dynamics occur for most initial conditions. The higher the field intensity, the more certainly the system will become entangled, regardless of the initial condition. Moreover, a sharp and monotonic growth of the entanglement was in general obtained for early time evolutions, reason why these type of novel molecular systems are suitable to be proposed for new quantum protocols.

\section*{Acknowledgements}

K.R. thanks Ian Picken and Christiane P. Koch for carefully reading the manuscript and their valuable comments. This work has been supported by Universidad del Valle under the internal project 7967. The authors acknowledge the support from the Colombian Science, Technology and Innovation Fundation--COLCIENCIAS ``Francisco Jos{\'e} de Caldas'' under project 110\-665842793 (contract FP-005-2015). We also thank the Center of Excellence for Novel Materials (CENM) at Universidad del Valle for financial support for the research group.

\bibliographystyle{ieeetr}
\bibliography{VOlaya_arXiv}

\end{document}